\documentstyle[twoside,fleqn,espcrc2,epsfig]{article}
\title{Zero-modes of the QED Neuberger Dirac operator\thanks{
This work was supported in part by the US Department of Energy under
contract DE-FG02-97ER41022 and by the Fonds zur F\"orderung der 
wissenschaftlichen Forschung under project P14435-TPH. Based on a
talk by BAB and a poster by RP.}}
\author{Bernd A. Berg\address{Department of Physics, The Florida State
    University, Tallahassee, FL~32306, USA }$^,$\address{John 
    von Neumann Institute for Computing, Forschungszentrum, 
    D-52425 J\"ulich, Germany},
  Urs M. Heller\address{School of Computational Science and Information 
  Technology,\\
  ~~The Florida State University, Tallahassee, FL 32306, USA},
  Harald Markum\address{Atominstitut, Technische
  Universit\"at Wien, A-1040 Vienna, Austria},
  Rainer Pullirsch$^{\rm d}$,
  and
  Wolfgang Sakuler$^{\rm d}$}
\begin{document}
\begin{abstract}
  We consider $4d$ compact lattice QED in the quenched approximation.
  First, we briefly summarize the spectrum of the staggered Dirac 
  operator and its connection with random matrix theory. Afterwards we 
  present results for the low-lying eigenmodes of the Neuberger 
  overlap-Dirac operator. In the strong coupling phase we find exact zero-modes.
  Subsequently we discuss possibly related topological excitations of 
  the U(1) lattice gauge theory.
\end{abstract}
\date{\today}
\maketitle

\section{Spectrum of the QED Dirac operator}

The spectrum of the QCD Dirac operator
\begin{equation} \label{Dirac_Op}
i D + im = \left( \matrix{ im & T \cr T^{\dagger} & im \cr}
\right)\ ~{\rm in\ a\ chiral\ basis}
\end{equation}
is related to universality classes of random matrix theory (RMT),
i.e. determined by the global symmetries of the QCD partition
function, for a review see~\cite{VeWe00}.  In RMT the matrix $T$ in
Eq.~(\ref{Dirac_Op}) is replaced by a random matrix with appropriate
symmetries, generating the chiral orthogonal (chOE), unitary (chUE),
and symplectic (chSE) ensemble, respectively. For SU(2)
and SU(3) gauge groups numerous results exist confirming the expected
relations.

We have investigated $4d$ U(1) gauge theory described by the action
$S \lbrace U_l \rbrace = \sum_p (1 - \cos \theta_p )$
with $U_l = U_{x,\mu} = \exp (i\theta_{x,\mu}) $ and
$
  \theta_p =
 \theta_{x,\mu} +
 \theta_{x+\hat{\mu},\nu} -
 \theta_{x+\hat{\nu},\mu} -
 \theta_{x,\nu}\ \ (\nu \ne \mu)\ . $
At $\beta_c \approx 1.01$ U(1) gauge theory undergoes a phase
transition between a confinement phase with mass gap and monopole
excitations for $\beta < \beta_c$ and the Coulomb phase for $\beta > \beta_c$.
In the whole Coulomb phase the photon is massless and thus there ought
to exist a continuum theory, as $V\to\infty$, everywhere.

We were interested in the relationship between U(1) gauge theory and
RMT across this phase transition. The
Bohigas-Giannoni-Schmit conjecture~\cite{BoGiSc84} states that quantum
systems whose classical counterparts are chaotic have spectral
fluctuation properties, measured, e.g. by the nearest-neighbor
spacing distribution $P(s)$ of unfolded eigenvalues, given by RMT,
whereas systems whose classical counterparts are integrable obey a
Poisson distribution, $P(s)=\exp (-s)$. 
With staggered fermions it was shown in Ref.~\cite{BeMaPu98} that the
Bohigas-Giannoni-Schmit conjecture holds for lattice QED both
in the confined phase (as for the SU(2) and SU(3) gauge 
groups) and also in the Coulomb phase, whereas free fermions yield the 
Poisson distribution. In Ref.~\cite{BeMaPu00} this investigation was 
continued with a study of the distribution of small eigenvalues in the 
confined phase. Excellent agreement was found for the conjecture that
the microscopic spectral density
\begin{equation} \label{rho_s}
 \rho_s (z) = \lim_{V\to\infty} {1\over V \Sigma}\,
 \rho \left( {z\over V\Sigma } \right)
\end{equation}
is given by the result for the chUE of RMT, which also generates the 
Leutwyler-Smilga sum rules~\cite{VeWe00}. Here $\Sigma$ is the modulus
of the chiral
condensate, which follows from the smallest non-zero eigenvalue via
the Banks-Casher formula.
\begin{figure}[ht]
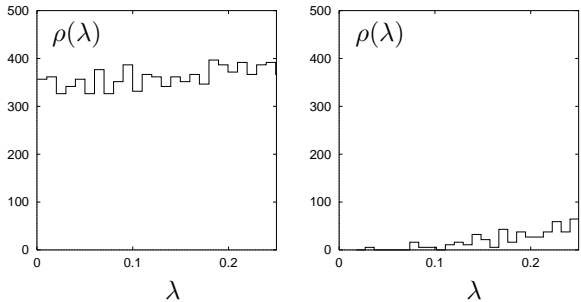

 \vspace{6mm}
 \hspace*{6mm}$\rho(\lambda)$\hspace*{33.7mm}$\rho(\lambda)$\vspace{-6mm}
 \centerline{\psfig{figure=f12.eps,width=3.5cm}\hspace*{4mm}
    \psfig{figure=f11.eps,width=3.5cm}}
  \vspace*{-2mm}
  \hspace*{21mm}$\lambda$\hspace*{38mm}$\lambda$
  \vspace*{-8mm}
  \caption{Density $\rho(\lambda)$ of small eigenvalues for the $8^3\times 6$ lattice
    at $\beta = 0.9$ (left plot) and at $\beta = 1.1$ (right plot). A
    non-zero chiral condensate is supported in the confinement phase.}
  \label{f12}
  \vspace*{-0.5cm}
\end{figure}
The quasi zero-modes which are responsible for the chiral condensate
build up when we cross from the Coulomb into the confined phase. For 
an $8^3\times 6$ lattice~\cite{BeMaPu00}, Fig.~\ref{f12} compares
on identical scales densities of the small eigenvalues at $\beta =
0.9 $ (left plot) and at $\beta = 1.1$ (right plot), averaged over
20 configurations. The quasi zero-modes in the left plot are 
responsible for the non-zero chiral condensate $\Sigma>0$ via the
Banks-Casher formula, whereas no such quasi zero-modes are found 
in the Coulomb phase. For $4d$ SU(2) and SU(3) gauge theories a general 
interpretation is to link zero-modes to the existence of instantons. As 
there are no conventional instantons in $4d$ U(1) gauge theory
(an analogous case exists 
in $3d$ QCD \cite{VeZa93b}), we~\cite{BeHeMa01} decided to study the physical 
origin of the U(1) quasi zero modes in more detail. 

\section{Exact zero-modes}

\begin{figure}[ht]
  \vspace{-2mm}
  \begin{center}
    \epsfig{figure=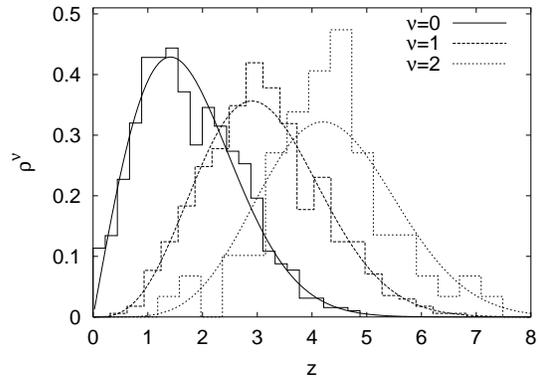,width=\columnwidth}
    \vspace{-15mm}
    \caption{Exact RMT probability densities~(\ref{rho_nu})
      for the lowest non-zero eigenvalues compared with
      histograms from our data.}
    \label{fig_pd}
  \end{center}
\vspace{-3mm}
\end{figure}
Via the Atiyah-Singer index theorem the topological charge is mapped 
on the number of fermionic zero-modes of the Dirac operator. It
is thus desirable to use a Dirac operator which retains 
chiral symmetry and exhibits exact zero-modes. Therefore, we employed
in Ref.~\cite{BeHeMa01} the Neuberger overlap-Dirac operator~\cite{Neu98,Nara95} 
\begin{equation} \label{Neuberger_Dirac}
D = {1\over 2}\, \left[ 1 + \gamma_5\, \epsilon 
\left( H_w(m) \right) \right]\ .
\end{equation}
Here $\gamma_5 H_w(-m)$ is the usual Wilson-Dirac operator on the 
lattice and $\epsilon$ the sign function. The mass $m$ has to be
chosen to be positive and well above the critical mass for Wilson
fermions but below the mass where the doublers become light on the
lattice.

Our first question was whether the QED overlap-Dirac operator will exhibit
exact zero-modes at all. To answer it, we have analyzed configurations on 
$L^4$ lattices at $\beta =0.9$ in the confined phase and at $\beta =1.1$ 
in the Coulomb phase. With an overrelaxation/heatbath algorithm we 
generated 500 configurations per lattice of linear size $L=4$, $6$ and 
$8$. After thermalization, the configurations were separated by 100 
sweeps, with each sweep consisting of three overrelaxation updates of each 
link, followed by one heatbath update. On each configuration the lowest 
12 eigenvalues of the overlap-Dirac operator~(\ref{Neuberger_Dirac})
were calculated using the Ritz functional algorithm with the optimal 
rational approximation of Ref.~\cite{EHN99} to 
$\epsilon \left( H_w(m) \right)$ and $m$ set to 2.3. 

\begin{figure*}[ht]
\centerline{\hspace*{16mm}
\psfig{figure=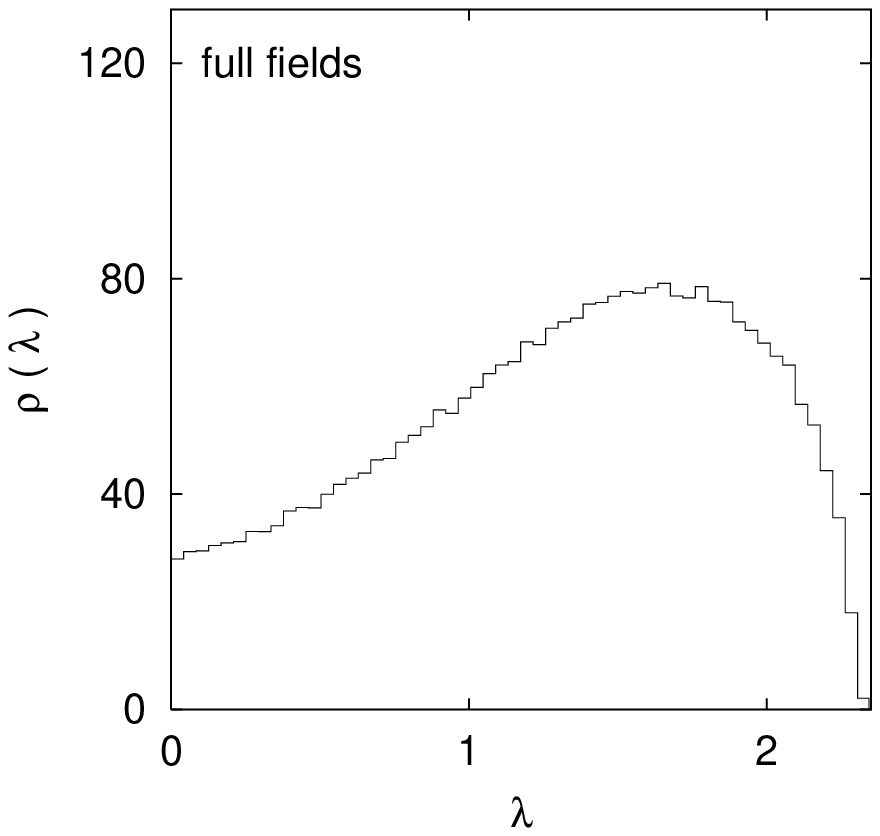,width=6.9cm,height=4cm}\hspace*{-15mm}
\psfig{figure=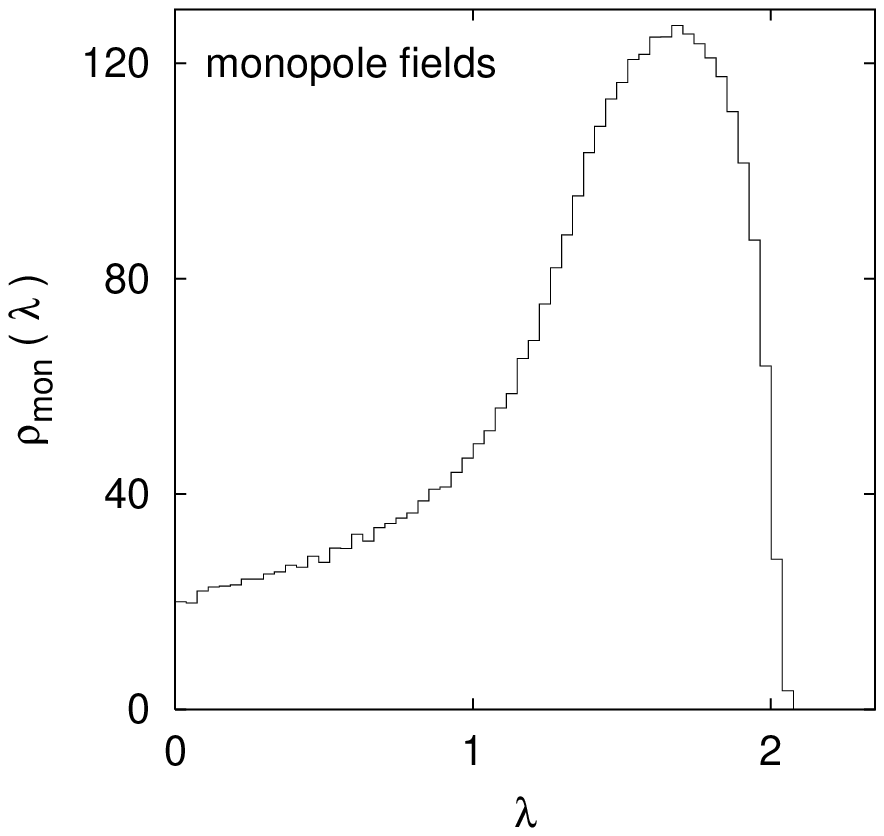,width=6.9cm,height=4cm}\hspace*{-15mm}
\psfig{figure=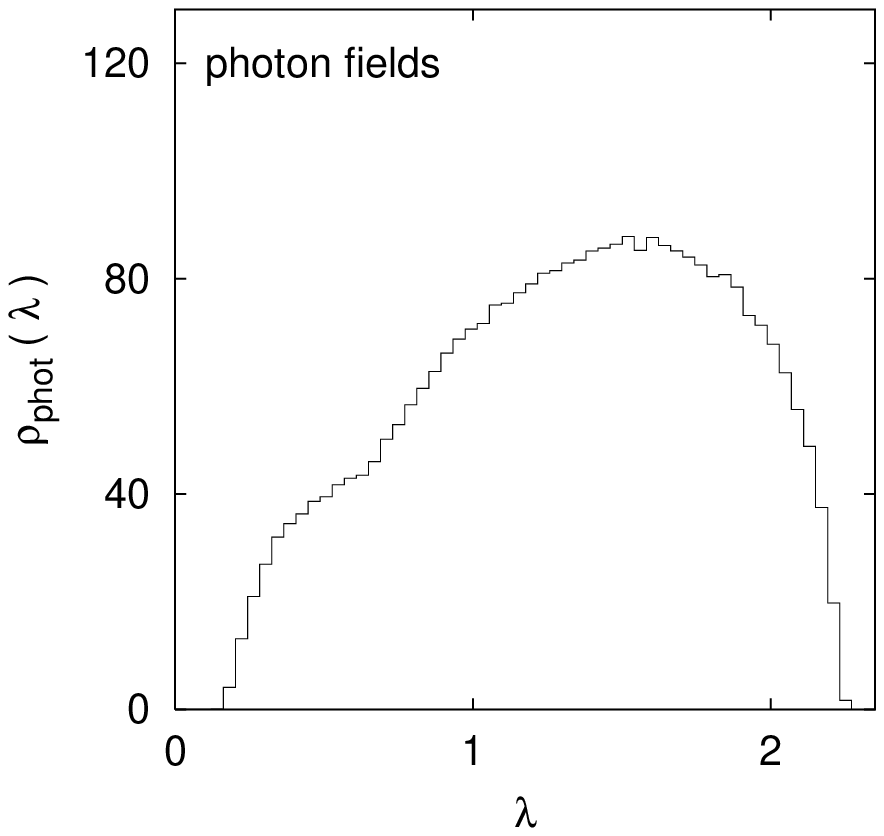,width=6.9cm,height=4cm\hspace*{-0mm}}}
\vspace*{-10mm}
\caption{Decomposition of the spectrum $\rho(\lambda)$ (normalized to the
         number of eigenvalues) of the staggered Dirac operator into a
         monopole part $\rho_{\rm mon}(\lambda)$ and a photon part
         $\rho_{\rm phot}(\lambda)$, averaged over 500 configurations
         on a $4^4$ lattice.} \label{staggered}
\end{figure*}

In the confined phase exact zero-modes of the 
operator~(\ref{Neuberger_Dirac}) were indeed found. The highest degeneracy 
observed was $\nu =3$. No zero-modes were found in the Coulomb phase.
For each degeneracy $\nu$, chiral RMT predicts the distribution of the 
lowest non-zero eigenvalue $\lambda_{\min}$ in terms of the rescaled 
variable
\begin{equation} \label{z_def}
z = \Sigma\,V\,\lambda_{\min}\ ,
\end{equation}
where $V$ is the volume of the system and $\Sigma$ the infinite volume
value of the chiral condensate $\langle\overline{\psi}\psi\rangle$
determined up to an overall wave function renormalization. For the U(1) 
gauge group the unitary ensemble applies and the RMT predictions for the
$\nu=0,1,2$ probability densities $\rho^{(\nu)}_{\min}(z)$ 
are~\cite{VeWe00}
\begin{equation} \label{rho_nu}
\rho^{(\nu)}_{\min}(z)\!=\!\cases{ \!{z\over 2}e^{-{z^2\over4}}(\nu=0)\cr
\!{z\over 2}I_2(z)\,e^{-{z^2\over 4}}(\nu=1)\cr
\!{z\over 2}[I_2(z)^2\!-\!I_1(z)I_3(z)]\,e^{-{z^2\over4}}(\nu=2).}
\end{equation}
The chiral condensate is related to the expectation value of 
the smallest eigenvalue. For degeneracy $\nu$ we have
\begin{equation} \label{Sigma}
 \Sigma\ =\ \Sigma^{(\nu)}\ =\ {\langle z^{(\nu)}\rangle \over V\, 
 \langle\lambda^{(\nu)}_{\min}\rangle} \ ,
\end{equation}
where the result is supposed to be independent of $\nu$ and
\begin{equation} \label{average_z}
 \langle z^{(\nu)} \rangle = \int_0^{\infty} dz\, z\, 
 \rho^{(\nu)}_{\min} (z)\ .
\end{equation}
Using analytical and numerical integration the expectation values 
$\langle z^{(\nu)} \rangle$ are easily calculated. The values 
$\Sigma^{(\nu)}$ follow then from Eq.~(\ref{Sigma}) and, for the 
various lattice sizes and $\nu$'s, the results which we obtain 
from our data are consistent with one another.

Figure~\ref{fig_pd} displays the exact RMT probability 
densities $\rho^{(\nu)}_{\min}$ of Eq.~(\ref{rho_nu}) and the 
corresponding histograms from our data. It
shows that the histograms follow the shift of the RMT probability
densities and their general shape. The high peak of the $\nu=2$
histogram is interpreted as a statistical fluctuation due to
the low statistics we have for this case.

\section{Topological objects}

The topological structure of the U(1) gauge theory is not as 
extensively studied as the non-Abelian case. Nevertheless, 
with torons~\cite{torons}, monopole solutions~\cite{mono} and Dirac
sheets~\cite{GJJ85} a number of topological objects are known for
the U(1) gauge theory.  All these topological
configurations have in common with the zero-mode degeneracy that
they are turned on at the phase transition from order (Coulomb)
to disorder (confinement). However, in the confined phase  we found
no detailed correlation between any of the topological phenomena
and the zero-mode degeneracy of the overlap-Dirac operator. 
Related to this may be that we found the zero-mode susceptibility
to decrease, possibly approaching zero for $L\to\infty$~\cite{BeHeMa01}.

In contrast to these findings there is an interesting
observation that ordered Dirac sheet configurations give rise to
zero-modes of the overlap-Dirac operator, with the number of zero-modes
being equal to the number of Dirac sheets~\cite{Nara95,Chiu99}.
One might understand this from the fact that a Dirac sheet is a
$2d$ gauge configuration that contains unit topological charge in
the $2d$ sense, kept constant in the two orthogonal directions.

To investigate possible correlations between the existence of zero-modes
and topological objects further we have, following
Refs.~\cite{StWe92,Suzu96,Biel97}, factorized our gauge configurations
into monopole and photon fields.
The U(1) plaquette angles $\theta_{x,\mu\nu}$ are decomposed into the
``physical'' electromagnetic flux through the plaquette 
$\bar \theta_{x,\mu\nu}$ and a number $m_{x,\mu\nu}$ of Dirac strings 
passing through the plaquette
\begin{equation} \label{Dirac_string_def}
 \theta_{x,\mu\nu} = \bar \theta_{x,\mu\nu} + 2\pi\,m_{x,\mu\nu}\ ,
\end{equation}
where $\bar \theta_{x,\mu\nu}\in (-\pi,+\pi]$ and
$m_{x,\mu\nu} \ne 0$ is called a Dirac plaquette. Monopole and photon
fields are then defined in the following way
\begin{equation} \label{monoplaq}
\theta^{\rm mon}_{x, \mu} = - 2 \pi\, \sum_{x'} G_{x,x'} \,
\partial_{\nu}' \, m_{x', \nu\mu}
\end{equation}
\begin{equation} \label{photplaq}
\theta^{\rm phot}_{x, \mu} = - \, \sum_{x'} G_{x,x'} \,
\partial_{\nu}' \, \bar\theta_{x', \nu\mu} \ .
\end{equation}
Here $\partial'$ acts on $x'$, the quantities
$m_{x,\mu\nu}$ and $\bar\theta_{x, \mu\nu}$ are defined in
Eq.~(\ref{Dirac_string_def}) and $G_{x,x'}$ is the lattice Coulomb 
propagator. One can show that $\tilde\theta_{x,\mu} \equiv
\theta^{\rm mon}_{x,\mu} +\theta^{\rm phot}_{x,\mu}$ is up to a gauge
transformation identical
with the original $\theta_{x,\mu}$ defined by $U_{x,\mu}=
\exp(i \theta_{x,\mu})$.

We found that both zero-modes and near zero-modes lie solely in the monopole 
part of the gauge field and are completely absent in the photonic 
field.  Using periodic boundary conditions in space and anti-periodic 
boundary conditions in time~\cite{Biel97}, this was seen both for the 
overlap-Dirac operator and for the quasi zero-modes of the staggered 
Dirac operator. As an example we show in Fig.~\ref{staggered} the entire
spectral density of the staggered Dirac operator in the original (full)
gauge fields and in their monopole and photon part, respectively. 

At the moment we are accumulating statistics to perform a decomposition
into different topological sectors and to obtain an analogous analysis
concerning topological objects as we did with the original U(1)
field above~\cite{BeHeMa01}. It is of further interest
to study space-time correlations between different topological objects.
We pose the question of the existence of local correlations between the
topological charge density and the monopole density. It is also desirable
to calculate spatial correlations between the topological objects and
the density $\psi^\dagger \psi(x)$ of the exact zero-modes of the 
overlap-Dirac operator.

\section{Cooling}
\begin{figure}[ht]
\vspace{-5mm}
\begin{center}
              \epsfig{figure=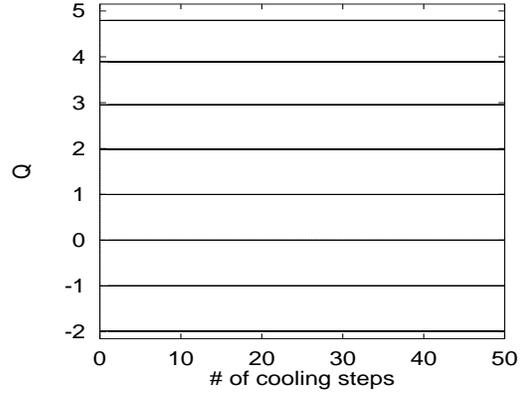,width=9.5cm,height=5.5cm}
\end{center}
\vspace*{-10mm}
    \caption{\label{artificial}
            Topological charge of artificially constructed configurations
             on an $8^4$ lattice.}
\end{figure}
\begin{figure}[ht]
\begin{center}
              \epsfig{figure=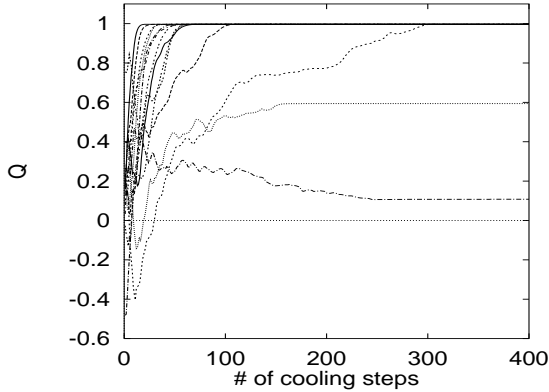,width=9.5cm,height=5.5cm}
\end{center}
\vspace*{-10mm}
    \caption{\label{heatcool}
             Cooling history of the topological charge of the $Q=1$ configuration
             after increasing number of updates at $\beta=0.9$.
             Integer topological charge is lost after $^{_<}_{^{\sim}} 10$ updates.}
\end{figure}
\begin{figure}[ht]
\begin{center}
              \epsfig{figure=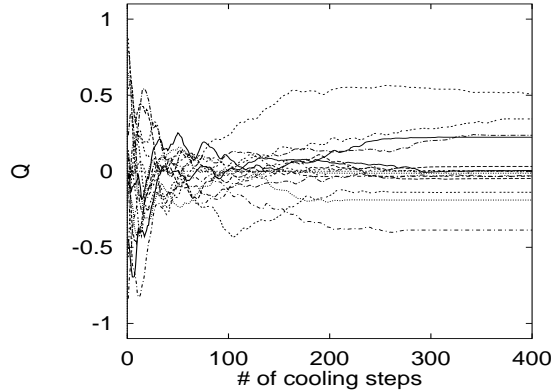,width=9.5cm,height=5.5cm}
\end{center}
\vspace*{-10mm}
    \caption{\label{realtopo}
             Cooling history of typical equilibrium configurations
             at $\beta=0.9$. No integer topological charge is found.}
\end{figure}

In non-Abelian theories it turned out that some cooling or smoothing
procedure is unavoidable to gain integer-valued topological charges.
For our Abelian case we studied the value $Q$ obtained by integration
of the topological density
\begin{equation}
 Q = \int d^4x \frac{1}{32\pi^2}F_{\mu\nu}(x) \tilde{F}_{\mu\nu}(x)
\end{equation}
using the hypercube definition and plaquette definition on the 
lattice~\cite{diVe81}.
In $4d$ U(1) theory with periodic boundary conditions one can
construct an artificial configuration with integer topological charge 
$Q$~\cite{SmVi87}
\begin{eqnarray}
U_{1}(x) &=& \exp( i \omega_{1} a x_{2})  \nonumber \\
U_{2}(x) &=& 1  \hspace*{9mm}            (x_{2} = a,2a,...,(N-1)a)  \nonumber\\
U_{2}(x) &=& \exp(-i \omega_{1} a N x_{1})  \hspace*{8.2mm}   (x_{2} = Na)  \nonumber\\
U_{3}(x) &=& \exp( i \omega_{2} a x_{4})  \nonumber\\
U_{4}(x) &=& 1  \hspace*{9mm}            (x_{4} = a,2a,...,(N-1)a)  \nonumber\\
U_{4}(x) &=& \exp(-i \omega_{2} a N x_{3})   \hspace*{8.2mm}  (x_{4} = Na),
\end{eqnarray}
where $N$ is the linear lattice extent,
$ \omega_{i} a^2 = \frac{2 \pi n_{i}}{N^2} $, $n_{i} = ...,-2,-1,0,1,2,...$,
and $Q = n_{1} n_{2}$. These gauge fields consist of $n_{1}$ Dirac sheets
in the 1-2 planes and $n_{2}$ Dirac sheets in the 3-4 planes.
Such configurations possess nearly integer-valued $Q$'s on a finite lattice,
as seen from Fig.~\ref{artificial}, and the overlap-Dirac operator has
$Q$ exact zero-modes.

We have first heated and then cooled those smooth configurations in order to
inspect whether they return to their original value of $Q$. In
Fig.~\ref{heatcool} we demonstrate the effect for the $Q=1$ configuration
which was updated several times with the Monte Carlo code at $\beta=0.9$
before 400 cooling steps were applied. We employed a systematic cooling
procedure by $U_{\mu}(x) \to VU_{\mu}(x)$, with $V=\frac{A^*}{|A|}$ and
$A=\sum\limits_{\eta=\pm 1}
\sum\limits_{\nu \neq \mu} U_{\mu}(x) S^{\eta}_{\mu\nu}(x)$.
It turned out that the artificial configuration has a memory of $Q=1$ for
up to 10 Monte Carlo hits. When cooling equilibrium configurations from
the $\beta=0.9$ ensemble it never happened that they obtained an
integer-valued topological charge, as depicted in Fig.~\ref{realtopo}.

\section{Summary}

\noindent
$\bullet$ Exact zero-modes of the overlap-Dirac operator are observed
	in the confined phase of compact U(1) gauge theory.\\
$\bullet$ Agreement of the distribution of the smallest non-zero eigenvalue
        with predictions from random-matrix theory for fixed zero-mode
	degeneracy is found.\\
$\bullet$ No obvious correlations between toron charge, number of monopoles,
	number of Dirac sheets and the number of exact zero-modes is seen.\\
$\bullet$ Exact zero-modes survive in the monopole part of decomposed gauge
	field configurations, but not in the photon part.\\
$\bullet$ No equilibrium configurations with integer topological charge are
	found. Cooling does not lead to integer values for compact U(1) gauge
	theory.\\
$\bullet$ Only artificially constructed configurations have integer topological
	charge. Only for $^{_<}_{^{\sim}} 10$ Monte Carlo updates can this
	charge be retained after cooling.\\
$\bullet$ Local spatial correlations between monopoles, topological charge
	density, and eigenfunctions of overlap-Dirac operator are analyzed
	currently to shed light on the confining mechanism of compact QED.\\

\end{document}